\documentstyle[epsfig]{aipproc}

\begin{document}
\title{A renormalisation-group treatment of two-body scattering}
\author{Michael C. Birse,  Judith A. McGovern and Keith G. Richardson}
\address{Theoretical Physics Group, Department of Physics and Astronomy\\
University of Manchester, Manchester, M13 9PL, UK\\}

\maketitle

\begin{abstract}
A Wilsonian renormalisation group is used to study nonrelativistic two-body 
scattering by a short-ranged potential. We identify two fixed points: a trivial 
one and one describing systems with a bound state at zero energy. The
eigenvalues of the linearised renormalisation group are used to assign a
systematic power-counting to terms in the potential near each of these fixed
points. The expansion around the nontrivial fixed point is shown to be 
equivalent to the effective-range expansion.
\end{abstract}

\section*{Introduction}

Recently there has been much interest in the possibility of developing a
systematic treatment of low-energy nucleon-nucleon scattering using the
techniques of effective field theory\cite{mcb:wein2,mcb:ksw2,mcb:vk3}. 
Here we approach the problem using Wilson's continuous renormalisation 
group\cite{mcb:wrg} to 
examine the low-energy scattering of nonrelativistic particles interacting
through short-range forces\cite{mcb:bmr}.

The starting point for the renormalisation group (RG) is the imposition of
a momentum cut-off, $|{\bf k}|<\Lambda$, separating the low-momentum physics 
which we are interested in from the high-momentum physics which we wish to 
``integrate out". Provided that there is a separation of scales between these 
two regimes, we may demand that low-momentum physics should be independent of 
$\Lambda$.

The second step is to rescale the theory, expressing all dimensioned quantities 
in units of $\Lambda$. As the cut-off $\Lambda$ approaches zero, all physics
is integrated out until only $\Lambda$ itself is left to set the scale. In 
units of $\Lambda$ any couplings that survive are just numbers, and these 
define a ``fixed point". Such fixed points correspond to systems with no
natural momentum scale. Examples include the trivial case of a zero scattering
amplitude and the more interesting one of a bound state at exactly zero energy.

Real systems can then be described in terms of perturbations away from one of
these fixed points. For perturbations that scale as definite powers of 
$\Lambda$, we can set up a power-counting scheme: a systematic way to organise 
the terms in an effective potential or an effective field theory. A fixed
point is said to be stable if all perturbations vanish like positive powers of 
$\Lambda$ as $\Lambda\rightarrow 0$ and unstable if one or more of them grows
with a negative power of $\Lambda$.

\section*{Two-body scattering}

We consider $s$-wave scattering by a potential that consists of contact 
interactions only. Expanded in powers of energy and momentum this has the form
\begin{equation} \label{mcbeq:pot1}
V(k',k,p)=C_{00}+C_{20}(k^2+k'^2)+C_{02}\,p^2\cdots,
\end{equation}
where $k$ and $k'$ denote momenta and energy-dependence is expressed in 
terms of the on-shell momentum $p=\sqrt{ME}$. Below all thresholds for 
production of other particles, this potential should be an analytic function of 
$k^2$, $k'^2$ and $p^2$.

Low-energy scattering is conveniently described in terms of the reactance 
matrix, $K$. This is similar to the scattering matrix $T$, except for the use 
of standing-wave boundary conditions. It satisfies the Lippmann-Schwinger (LS) 
equation (see\cite{mcb:newt})
\begin{equation} \label{mcbeq:lse}
K(k',k,p)=V(k',k,p)+\frac{M}{2\pi^2}{\cal{P}}
\int q^2dq\,\frac{V(k',q,p)K(q,k,p)}
{{p^2}-{q^2}},
\end{equation}
where ${\cal P}$ denotes the principal value.

On-shell, with $k=k'=p$, the $K$-matrix is related to the phase-shift by
\begin{equation}
\frac{1}{K(p,p,p)}=-\frac{M}{4\pi}p\cot\delta(p),
\end{equation}
which means it has a simple relation to the effective-range 
expansion\cite{mcb:ere}, 
\begin{equation} \label{mcbeq:ere}
p\cot\delta(p)-\frac{1}{a}+\frac{1}{2}r_{e}p^2+\cdots,
\end{equation}
where $a$ is the scattering length and $r_e$ is the effective range. We shall
see that this turns out to be equivalent to an expansion around a nontrivial
fixed point of the RG.

\section*{Renormalisation group}

To set up the RG we first impose a momentum cut-off on the intermediate states 
in the LS equation (\ref{mcbeq:lse}). This can be written
\begin{equation}\label{mcbeq:slse}
K=V(\Lambda)+V(\Lambda)G_0(\Lambda)K,
\end{equation}
where we have included a sharp cut-off in the free Green's function,
\begin{equation}
G_0={M \theta(\Lambda-q)\over p^2-q^2}.
\end{equation}

We now demand that $V(k',k,p,\Lambda)$ varies with $\Lambda$ in order to
keep the off-shell $K$-matrix independent of $\Lambda$:
\begin{equation}
{\partial K\over\partial\Lambda}=0.
\end{equation}
This is sufficient to ensure that all scattering observables do not depend on 
$\Lambda$. Differentiating the LS equation (\ref{mcbeq:slse}) with respect to 
$\Lambda$ and then operating from the right with $(1+G_0K)^{-1}$, we get
\begin{equation}\label{mcbeq:rge}
{\partial V\over\partial\Lambda} 
={M\over2\pi^2}V(k',\Lambda,p,\Lambda){\Lambda^2\over\Lambda^2-p^2}
V(\Lambda,k,p,\Lambda).
\end{equation}

We now introduce dimensionless momentum variables, $\hat k=k/\Lambda$ etc.,
and a rescaled potential,
\begin{equation}
\hat V(\hat k',\hat k,\hat p,\Lambda)={M\Lambda\over 2\pi^2}
V(\Lambda\hat k',\Lambda\hat k,\Lambda\hat p,\Lambda).
\end{equation}
From the equation (\ref{mcbeq:rge}) satisfied by $V$ we find that the rescaled
potential satisfies the RG equation 
\begin{equation}\label{mcbeq:scrge}
\Lambda{\partial\hat V\over\partial\Lambda}
=\hat k'{\partial\hat V\over\partial\hat k'}
+\hat k{\partial\hat V\over\partial\hat k}
+\hat p{\partial\hat V\over\partial\hat p}
+\hat V
+\hat V(\hat k',1,\hat p,\Lambda){1\over 1-\hat p^2}
\hat V(1,\hat k,\hat p,\Lambda).
\end{equation}

\section*{Fixed points}

We are now in a position to look for fixed points: solutions of 
(\ref{mcbeq:scrge}) that are independent of $\Lambda$. These provide the 
possible low-energy limits of theories as $\Lambda\rightarrow 0$ and hence the 
starting points for systematic expansions of the potential.

\subsection*{The trivial fixed point}

One obvious solution of (\ref{mcbeq:scrge}) is the trivial fixed point, 
\begin{equation}
\hat V(\hat k',\hat k,\hat p,\Lambda)=0,
\end{equation} 
which describes a system with no scattering.

For systems described by potentials close to the fixed point we can expand in 
terms of eigenfunctions, $\hat V=\Lambda^\nu\phi(\hat k',\hat k,\hat p)$, of 
the linearised RG equation,
\begin{equation}\label{mcbeq:linrge.tr}
\hat k'{\partial\phi\over\partial\hat k'}
+\hat k{\partial\phi\over\partial\hat k}
+\hat p{\partial\phi\over\partial\hat p}
+\phi=\nu\phi.
\end{equation}
These have the form
\begin{equation}
\hat V(\hat k',\hat k,\hat p,\Lambda)=C\Lambda^\nu 
\hat k^{\prime l}\hat k^m \hat p^n,
\end{equation}
with eigenvalues $\nu=l+m+n+1$, where $l$, $m$ and $n$ are non-negative even
integers. The eigenvalues are all positive and so the fixed point is a
stable one: all nearby potentials flow towards it as $\Lambda\rightarrow 0$.

The corresponding unscaled potential has the expansion
\begin{equation}\label{mcbeq:potexp.tr}
V(\hat k',\hat k,\hat p,\Lambda)={2\pi^2\over M}\sum_{l,n,m}\widehat C_{lmn}
\Lambda_0^{-\nu} k^{\prime l} k^m p^n,
\end{equation}
where we have written the coefficients in dimensionless form by taking out 
powers of $\Lambda_0$, the scale of the short-distance physics. The power 
counting in this expansion is just the one proposed by Weinberg\cite{mcb:wein2} 
if we assign an order $d=\nu-1$ to each term in the potential. This fixed point 
can be used to describe systems where the scattering at low energies is weak 
and can be treated perturbatively. It is not the appropriate starting point for 
$s$-wave nucleon-nucleon scattering, where the scattering length is large.

\subsection*{A nontrivial fixed point}

The simplest nontrivial fixed point is one that depends on energy only,
$\hat V=\hat V_0(\hat p)$. It satisfies
\begin{equation}\label{mcbeq:fprge.ere}
\hat p{\partial\hat V_0\over\partial\hat p}
+\hat V_0(\hat p)+{\hat V_0(\hat p)^2\over 1-\hat p^2}=0.
\end{equation}
The solution, which must be analytic in $\hat p^2$, is
\begin{equation}
\hat V_0(\hat p)=-\left[1-{\hat p\over 2}\ln{1+\hat p\over 1-\hat p}
\right]^{-1}.
\end{equation}
Although the detailed form of this potential is specific to our particular 
choice of cut-off, the fact that it tends to a constant as $\hat p\rightarrow 
0$ is a generic feature, which is present for any regulator. 

The corresponding unscaled potential is
\begin{equation}\label{mcbeq:potfp.ere}
V_0(p,\Lambda)=-{2\pi^2\over M}\left[\Lambda-{p\over 2}
\ln{\Lambda+p\over\Lambda-p}\right]^{-1}.
\end{equation}
The solution to the LS equation for $K$ with this potential is
infinite, or rather $1/K=0$. This corresponds to a system with infinite
scattering length, or equivalently a bound state at exactly zero energy.

To study the behaviour near this fixed point we consider small perturbations
about it that scale with definite powers of $\Lambda$: 
\begin{equation}
\hat V(\hat k',\hat k,\hat p,\Lambda)=\hat V_0(\hat p)
+C\Lambda^\nu \phi(\hat k',\hat k,\hat p).
\end{equation}
These satisfy the linearised RG equation
\begin{equation}\label{mcbeq:linrge.ere}
\hat k'{\partial\phi\over\partial\hat k'}
+\hat k{\partial\phi\over\partial\hat k}
+\hat p{\partial\phi\over\partial\hat p}+\phi
+{\hat V_0(\hat p)\over 1-\hat p^2}\left[
\phi(\hat k',1,\hat p)+\phi(1,\hat k,\hat p)\right]
=\nu\phi.
\end{equation}

Solutions to (\ref{mcbeq:linrge.ere}) that depend only on energy ($\hat p$) 
can be found straightforwardly by integrating the equation. They are
\begin{equation}\label{enper.ere}
\phi(\hat p)=\hat p^{\nu+1} \hat V_0(\hat p)^2.
\end{equation}
Requiring that these be well-behaved as $\hat p^2\rightarrow 0$, we find the 
RG eigenvalues $\nu=-1,1,3,\dots$. The fixed point is unstable: it has one 
negative eigenvalue. 

\begin{figure}[h,t,b,p]
\vspace{0.5cm} \epsfysize=11cm \centerline{\epsffile{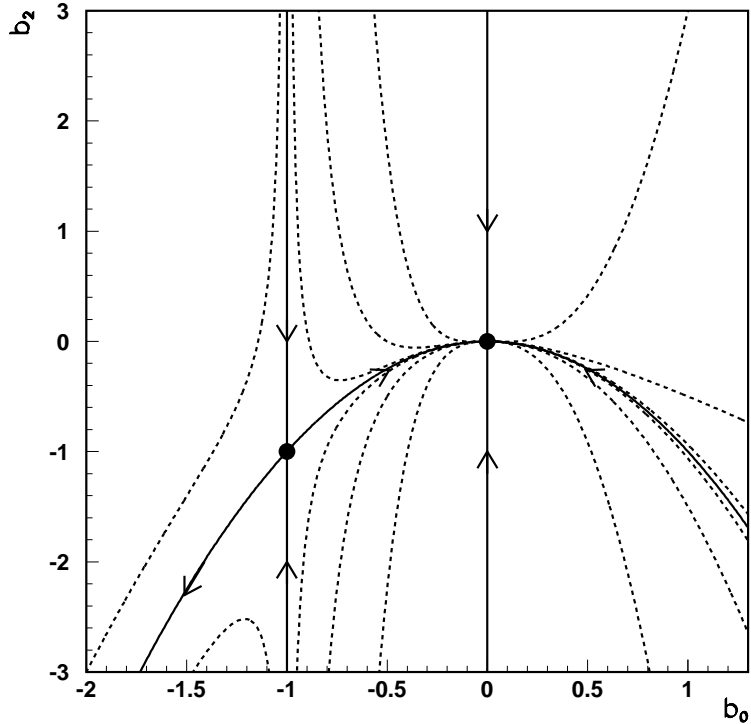}}
\centerline{{\caption{\label{fig:rgflow} The RG flow
of the first two terms in the expansion of the rescaled potential in powers of 
energy. The two fixed points are indicated by the black dots. The solid lines
are flow lines that approach one of the fixed points along a direction 
corresponding to an RG eigenfunction; the dashed lines are more general
flow lines. The arrows indicate the direction of flow as $\Lambda\rightarrow
0$. }}}
\end{figure}

The instability can be seen from the RG flow in Fig.~1. Only potentials
that lie exactly on the ``critical surface" flow into the nontrivial fixed 
point as $\Lambda\rightarrow 0$. Any small perturbation away from this 
surface eventually builds up and drives the potential either to the trivial 
fixed point at the origin or to infinity.

The corresponding unscaled potential is
\begin{equation}\label{mcbeq:pot.ere}
V(k',k,p,\Lambda)
=V_0(p,\Lambda)+{M\over 2\pi^2}\left(C_{-1}+C_1 p^2+\cdots\right)
V_0(p,\Lambda)^2.
\end{equation}
For perturbations around the nontrivial fixed point, we can assign an order 
$d=\nu-1=-2,0,2,\dots$ to each term in the potential. This power counting for 
(energy-dependent) perturbations agrees with that found by Kaplan, Savage and
Wise\cite{mcb:ksw2} using a ``power divergence subtraction" scheme and also by
van Kolck\cite{mcb:vk3} in a more general subtractive renormalisation scheme. 
The equivalence can be seen by making the replacement
\begin{equation}
V_0=-{2\pi^2\over M\Lambda}+\cdots\rightarrow -{4\pi\over M\mu},
\end{equation}
where $\mu$ is the renomalisation scale introduced by Kaplan, Savage and 
Wise in their subtraction scheme, and which plays an analogous role to the 
cut-off $\Lambda$ in our approach.

The on-shell $K$-matrix for this potential is (to any order in the $C$'s)
\begin{equation}\label{mcbeq:enere.ere}
\frac{1}{K(p,p,p)}=-{M\over 2\pi^2}\left(C_{-1}+C_1 p^2+\cdots\right).
\end{equation}
This is just the effective-range expansion (\ref{mcbeq:ere}). There is a 
one-to-one correspondence between the perturbations in $V$ and the terms in 
that expansion,
\begin{equation}
C_{-1}=-{\pi\over 2a},\qquad 
C_1={\pi r_e\over 4}.
\end{equation}

The expansion around the nontrivial fixed point is the relevant one
for systems with large scattering lengths, such as $s$-wave nucleon-nucleon
scattering.

\section*{Weak long-range forces}

The treatment outlined above is only valid at very low momenta, where all
pieces of the potential can be regarded as short-range. To extend it to 
describe nucleon-nucleon scattering at higher momenta, we would like to include 
pion-exchange forces explicitly. The longest-ranged of these is single pion
exchange, which provides a central Yukawa potential, 
\begin{equation}
V_{1\pi}({\bf k}',{\bf k})=-{4\pi\alpha_\pi\over({\bf k}-{\bf k}')^2+m_\pi^2},
\end{equation}
where 
\begin{equation}
\alpha_\pi={g_A^2m_\pi^2\over 16\pi f_\pi^2}\simeq 0.072.
\end{equation}

As in chiral perturbation theory, we want to treat the pion mass as a new
low-energy scale (in addition to the momentum and energy variables). This 
can be done by defining a rescaled variable $\hat m_\pi=m_\pi/\Lambda$ and
applying the RG as above. The corresponding term in the rescaled potential
is
\begin{equation}
\hat V_{1\pi}(\hat{\bf k}',\hat{\bf k},\hat m_\pi,\Lambda)
=-\Lambda {M g_A^2\over 8\pi^2 f_\pi^2}\,
{\hat m_\pi^2\over(\hat{\bf k}-\hat{\bf k}')^2+\hat m_\pi^2}.
\end{equation}
It scales as $\Lambda^1$, like the effective-range term in the potential
above. This suggests that one-pion exchange (OPE) can be treated as a
perturbation. It would contribute at next-to-leading order (NLO) in the
potential.

However questions remain about whether OPE is really weak enough for a
perturbative treatment to be useful. A possible scale for nonperturbative
long-range physics is the pionic ``Bohr radius":
\begin{equation}
R={2\over\alpha_\pi M}\simeq 5.8\ \hbox{fm}.
\end{equation}
This should be compared with the range of the Yukawa potential, $r_\pi
=1/m_\pi=1.4$ fm, which cuts off the potential at long distances, preventing
the formation of a bound state. The ratio of these scales is
\begin{equation}
{r_\pi\over R}\simeq  0.24,
\end{equation}
Although this is smaller than the critical value of 0.84, at which a bound 
state forms\cite{mcb:newt}, one might expect relatively slow convergence of the 
perturbation series.

Further questions are raised when the contribution of OPE to the effective
range is examined. A perturbative treatment (to NLO in an expansion in powers 
of momenta, $m_\pi$ and $1/a$, as in\cite{mcb:ch}) gives a short-range 
contribution to the effective $^1S_0$ range of
\begin{eqnarray}\label{mcbeq:pcere}
r_e^0&=&r_e-{2\alpha_\pi M\over m_\pi^2}\\
&=&2.62-1.38=1.24\ \hbox{fm}.
\end{eqnarray}
It is also possible to set up a distorted-wave effective-range expansion, in
which the long-range interaction is treated all orders\cite{mcb:hk}. This is 
essentially an expansion in powers of energy of 
$p\cot(\delta-\delta_{1\pi})/|{\cal F}_{1\pi}(p)|^2$
where $\delta_{1\pi}$ is the OPE phase shift and ${\cal F}_{1\pi}(p)$ the 
corresponding Jost function\cite{mcb:newt}. The resulting
purely short-range effective range is\cite{mcb:r99} (see also\cite{mcb:sf})
\begin{equation}
r_e^0=4.2\ \hbox{fm}.
\end{equation}
This is significantly different from the perturbatively corrected 
effective range (\ref{mcbeq:pcere}). The difference may be an indication of 
either strong forces with two-pion range, or of strong short-range forces with 
a complicated structure\cite{mcb:ks}.

\section*{Summary}

We have applied Wilson's renormalisation group to nonrelativistic two-body
scattering and identified two important fixed points\cite{mcb:bmr}. 

The first is the trivial fixed point. Perturbations around it can be used to 
describe systems with weak scattering. These perturbations can be organised 
according to Weinberg's power counting\cite{mcb:wein2}.

The second fixed point describes systems with a bound state at exactly zero
energy. In this case the relevant power-counting is the one found by Kaplan, 
Savage and Wise\cite{mcb:ksw2} and van Kolck\cite{mcb:vk3}. The expansion 
around this fixed point is exactly equivalent to the effective-range expansion.

These ideas can be extended in various ways. Short-range interactions in other
numbers of spatial dimensions can be studied. The critical dimension for 
instability of the nontrivial fixed point is $D=2$, which has been studied for 
some time in the context of anyons\cite{mcb:j,mcb:mt}.

Three-body systems are also being studied from the point of view of effective 
field theory\cite{mcb:bhvk1,mcb:bg}. In some cases these display much more 
complicated behaviour under the RG than the two-body ones discussed 
above\cite{mcb:bhvk2}.

Various nucleon-nucleon scattering observables as well as deuteron properties
have been calculated using the expansion around the nontrivial fixed 
point\cite{mcb:ksw2,mcb:cgss,mcb:ssw,mcb:fms}. In this approach, pion-exchange 
forces are 
treated as perturbations. An alternative approach which is being explored by 
other groups is to use Weinberg's power counting in the expansion of the 
potential, but then to iterate that potential to all orders in the LS 
equation\cite{mcb:orvk,mcb:egm,mcb:vkrev,mcb:bmpvk,mcb:pc}. This may provide a 
way to evade the problems of slow convergence when OPE is included 
explicitly\cite{mcb:ch,mcb:ks}. 

Finally, strong long-ranged interactions, such as the Coulomb force, lead to
quite different behaviour from the examples discussed here. They can still be
treated using similar techniques, as in NRQED\cite{mcb:nrqed} and 
NRQCD\cite{mcb:nrqcd}.


\end{document}